\documentclass[aps,prl,showpacs,twocolumn,superscriptaddress]{revtex4}
\usepackage{graphicx}
\usepackage[latin1]{inputenc}
\usepackage{textcomp}
\usepackage{mathptmx}
\begin{document}

\title{Controlling the dynamics of an open many-body quantum system with localized dissipation}

\author{G. Barontini}
\email{barontini@physik.uni-kl.de}
\affiliation{Research Center OPTIMAS and Fachbereich Physik,
Technische Universit\"{a}t Kaiserslautern, 67663 Kaiserslautern, Germany}

\author{R. Labouvie}%
\affiliation{Research Center OPTIMAS and Fachbereich Physik,
Technische Universit\"{a}t Kaiserslautern, 67663 Kaiserslautern, Germany}

\author{F. Stubenrauch}%
\affiliation{Research Center OPTIMAS and Fachbereich Physik,
Technische Universit\"{a}t Kaiserslautern, 67663 Kaiserslautern, Germany}%

\author{A. Vogler}%
\affiliation{Research Center OPTIMAS and Fachbereich Physik,
Technische Universit\"{a}t Kaiserslautern, 67663 Kaiserslautern, Germany}%

\author{V. Guarrera}%
\affiliation{Research Center OPTIMAS and Fachbereich Physik,
Technische Universit\"{a}t Kaiserslautern, 67663 Kaiserslautern, Germany}

\author{H. Ott}
\affiliation{Research Center OPTIMAS and Fachbereich Physik,
Technische Universit\"{a}t Kaiserslautern, 67663 Kaiserslautern, Germany}%

\date{\today}

\begin{abstract}
We experimentally investigate the action of a localized dissipative potential on a macroscopic matter wave, which we implement by shining an electron beam on an atomic Bose-Einstein condensate (BEC). We measure the losses induced by the dissipative potential as a function of the dissipation strength observing a paradoxical behavior when the strength of the dissipation exceeds a critical limit: for an increase of the dissipation rate the number of atoms lost from the BEC becomes lower. We repeat the experiment for different parameters of the electron beam and we compare our results with a simple theoretical model, finding excellent agreement. By monitoring the dynamics induced by the dissipative defect we identify the mechanisms which are responsible for the observed paradoxical behavior. We finally demonstrate the link between our dissipative dynamics and the measurement of the density distribution of the BEC allowing for a generalized definition of the Zeno effect. Due to the high degree of control on every parameter, our system is a promising candidate for the engineering of fully governable open quantum systems.
\end{abstract}


\maketitle 

Gathering information from a quantum system is never free of cost. Every measurement process provides a coupling between the quantum system and the (classical) environment, which leads to non-unitary dynamics, and in some cases to the destruction of essentially quantum effects. The elusive transition from the quantum to the classical realm must therefore be inherent in the processes that the environment induces on the system. In recent decades several advances have been made in the study of environmentally induced phenomena like decoherence and decoherence-induced selection of preferred states (einselection) \cite{zurek,omnes,book}. More recently environmental action has been used to manipulate qubits in a system of trapped ions \cite{blatt}. The knowledge and the mastering of the action of the environment are essential for taming errors in quantum computation schemes \cite{shor,cory} or to engineer decoherence-free subspaces for qubits \cite{cirac,lidar,beige}, and are also key to understanding the emergence of classicality from the quantum \cite{omnes,book}. In the context of the theory of open quantum systems, environmental action gives rise to effective Hamiltonians which can contain imaginary terms \cite{omnes,book,ober}. Since these terms actually arise from a collection of an enormous number of degrees of freedom \cite{book}, however, a complete experimental control over them appears overly challenging. Here, we report the engineering of a fully controllable, environmentally induced imaginary potential acting on a quantum system, and present observations of the subsequent induced dynamics. The \emph{localized} imaginary potential is realized by the almost pure dissipative action of an electron beam (EB) on an atomic Bose-Einstein condensate (BEC). We show that such a potential can be used to describe a continuous measurement process that can exhibit a generalized version of the so-called Zeno effect. The combination of the robust and macroscopic many-body quantum behavior of a BEC and the high tunability and precision of the EB promotes such a system as a paradigm for governable open quantum systems. 

One of the most striking properties of BECs is that, despite their many-body nature, they can be described to a good approximation by a mean-field wavefunction obeying the so-called Gross-Pitaevskii equation (GPE). This remains valid also when the BEC is coupled with the environment. Starting from the Lindblad master equation $i\hbar \partial_t\hat{\rho} = [\hat{H},\hat{\rho}]+ i\hbar\hat{\mathcal{L}}\hat{\rho}$, where $\hat{\rho}$ is the density operator of the many-body system, $\hat{H}$ is the Hamiltonian operator, and $\hat{\mathcal{L}}$ is the dissipation operator such that $\hat{\mathcal{L}}\hat{\rho}=-\int d\textbf{x}\gamma(\textbf{x})/2
[\hat{\Psi}^+\hat{\Psi}\hat{\rho}+\hat{\rho}\hat{\Psi}^+\hat{\Psi}-2\hat{\Psi}\hat{\rho}\hat{\Psi}^+]$, with $\gamma(\textbf{x})$ the local dissipation rate, we can write the equation of motion for the expectation value of the bosonic field operator $\hat{\Psi}$ as $\partial_t\langle\hat{\Psi}\rangle=Tr(\hat{\Psi}d_t\hat{\rho})$, which leads to a time-dependent GPE with an additional imaginary term \cite{supp,brazy,sandro}:
\begin{equation}
i\hbar\frac{\partial\psi(\textbf{x},t)}{\partial t}=\left(-\frac{\hbar^2\nabla^2}{2m}+V_{ext}+g|\psi(\textbf{x},t)|^2-i\hbar\frac{\gamma(\textbf{x})}{2}\right)\psi(\textbf{x},t).
\label{dGPE}
\end{equation} 
Here $\psi$ is the BEC wavefunction, obeying the constraint $\int|\psi(\textbf{x},t)|^2d\textbf{x}=N(t)$, $V_{ext}$ the trapping potential and $g=4\pi\hbar^2a/m$, $a$ being the s-wave scattering length. Notably our technique allows independent control of the Hamiltonian and dissipative terms of this equation. The ability to describe our open quantum many-body system with such a simple expression is a key asset for understanding and mastering its dynamics. 

\paragraph*{Experimental implementation}
In our experiment we prepare a pure BEC of 75$\times10^3$ atoms in a single-beam optical trap by means of forced evaporation. Once the evaporation is over, we shine a focussed EB right at the center of the BEC. The EB is produced by a commercial electron microscope mounted inside the vacuum chamber \cite{gericke}. The electron microscope is able to generate a beam of 6 keV electrons with variable beam extensions and currents. When the electrons impact on the BEC they collide locally with the atoms, ionizing or exciting them. Those atoms which have undergone an electron collision escape from the trapping potential (Fig. 1a). The EB thus locally dissipates the BEC. The ionized atoms, roughly 40$\%$ of all those scattered, are then directed to an ion detector, where their arrival times are registered. While escaping from the trapping region the ions can collide with the trapped atoms producing additional losses \cite{supp}. The total detection efficiency $\eta$ is the product of the branching ratio (40\%) and of the combined ion optics and detector efficiency (75\%). Details of the experimental apparatus can be found in \cite{gericke,electronbeam}. If the EB is rapidly moved in a controlled pattern, the whole column density profile of the BEC can be reconstructed \cite{gericke,Peter}. Here we keep the EB fixed in the center of the BEC, and monitor the subsequent induced dynamics by looking at the temporal signal from the ion detector (Fig. 1b).  By controlling the beam parameters, we can engineer the dissipative term in equation (\ref{dGPE}): we write it as $\gamma(\textbf{x})=I\sigma/(2\pi e w^2)\exp(-(x^2+y^2)/2w^2)$, $I$ being the EB current, $\sigma$ the electron-BEC scattering cross-section \cite{supp}, $e$ the elementary charge and $w$ the standard deviation of the spatial electron distribution, assumed to be gaussian \cite{supp}.

\begin{figure}
	\centering
		\includegraphics[scale=0.48]{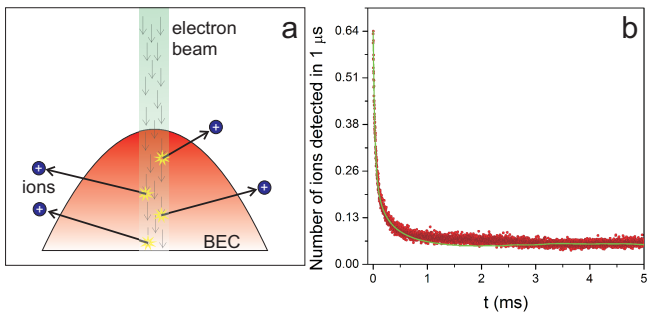}
	\caption{(Color online) a) The electrons locally collide with the atoms constantly dissipating the BEC. b) Temporal resolved signal from the ion detector. The bin size is 1 $\mu$s. Points are experimental data averaged over 1800 experimental repetitions, while the solid curve is the numerical simulation (see text). After 5 ms we typically collect $\simeq450$ ions.}
\label{Iw}
\end{figure}

\paragraph*{Comparison between experimental results and theoretical expectations}
In Fig. 2 we report the number of ions collected in the first 5 ms of continuous dissipation as a function of the EB current for three different values of $w$.  Notably we observe that the number of ions produced, as a function of the EB current (i.e., of the number of electrons sent on the atoms), shows a \emph{non-monotonic} dependence. In other words, starting from a critical value of the EB current, the harder we try to dissipate, the less we manage to do it. This paradoxical behavior is more marked for smaller values of $w$. In the same figure the data are compared with the results obtained by numerically solving equation (\ref{dGPE}), additionally taking into account secondary effects like ion-atom collisions \cite{supp}.  
The agreement  is very good (and the same agreement is visible in Fig. 1b), demonstrating that the description of the EB as a pure dissipative potential is sufficient to capture the observed main features. A detailed description of the dynamics that leads to the curves reported in Fig. 2 will be given in the following.  From simple textbook calculations, or from more formal analysis like the one made in \cite{allcock}, it is easy to verify that a \emph{localized} imaginary potential $U$ induces total reflection as the strength of the potential goes to infinity. Hence the effective quantum dissipation vanishes when the localized imaginary potential is either zero or infinity, implying the existence of a maximum of dissipation for some finite value of $U$. This explains on a qualitative basis the observed non-monotonicity. The position of the maximum is of special importance, since it sets the parameters which allow one to engineer the most efficient possible absorbing potential. As an example, in \cite{allcock}, where the time of arrival of a 1d wave packet is measured by a step-like potential, the maximum dissipation is analytically calculated to be $U_M\simeq10.6E$, where $E$ is the energy of the wave packet. In our case the presence of the non-linearity and the less idealized conditions do not allow for an analytic solution, but from the numerical results plotted in Fig. 3a we have found $U_M\simeq 8\mu \exp(-w/d)$, where $\mu$ and $d$ are respectively the chemical potential and the healing length of the unperturbed BEC. From Fig. 3a it also appears that increasing the size of the EB not only moves the position of the maximum dissipation to higher values of $I$, but also increases the number of produced ions for a given current, and reduces or washes out the effect of the reflection. Clearly, when $U=\hbar\gamma(\textbf{0})/2>U_M$, a decrease of the probe size (an increase of the resolution) leads to a lower production of ions, making the system more resistant to the environmental action.

\begin{figure}
	\centering
	\includegraphics[width=8.5cm]{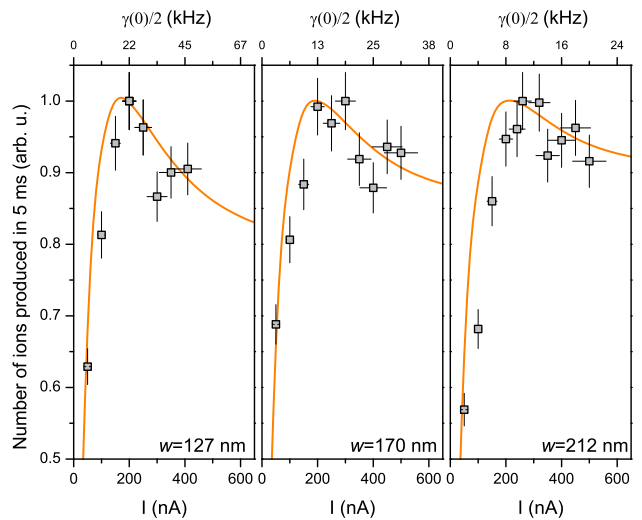} 
\caption{(Color online) Number of ions collected within the first 5 ms of continuous dissipation on a BEC as a function of the EB current $I$. The three panels report the data obtained with $w= 127 (5), 170 (7)$ and 212 (8) nm, from left to right. Each data point is the average over 75 experimental repetitions. The error bars are mainly due to shot-to-shot fluctuations in the overall ion detection efficiency. The solid lines are the number of dissipated atoms resulting from the numerical simulations (see text). Please note that in the sum  also the initial decay visible in Fig. 1b is included. The scale on the top reports the strength of the imaginary potential $U/\hbar=\gamma(\textbf{0})/2$ corresponding to the measured current.}
\label{figdiss}
\end{figure}

\paragraph*{Comparison to classical systems}
To ascertain to which extent our observations are peculiar to the wave nature of the BEC, we have repeated the experiment on a thermal gas of 4$\times10^5$ atoms at 1$\mu$K with a beam of $w=170$ nm. The results are reported in Fig. 3b, where a simple monotonic behavior is observed. Such data are well reproduced by a classical molecular dynamics simulation, which includes the dissipation induced by the EB \cite{supp}. 
We then extend the classical simulation to an atomic cloud which has the same density, number of atoms and trapping frequency as our BEC. Even though this does not represent any real physical system, it is instructive to compare the behavior of a quantum system with its hypothetical classical analogue. This comparison is made in Fig. 3c, where the monotonicity of the classical case is confirmed. From this we can conclude that the observed non-monotonicity is a purely quantum effect stemming from the \emph{macroscopic} wave nature of the BEC \footnote{Note that the non-monotonicity could be observed also for thermal particles provided that their wavefunction is significantly larger than the size of the dissipative potential.}. Moreover it is evident that the effect of the quantum reflection from the imaginary potential leads to a suppression of dissipation in the quantum case, which is already notable for very low currents.

\begin{figure}
\centering
	\includegraphics[width=8.5cm]{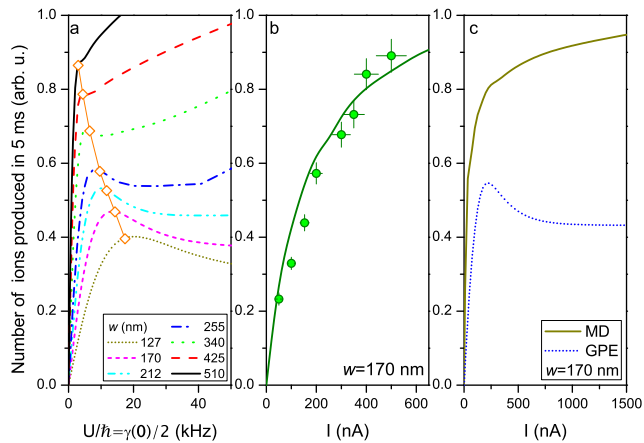}
	\caption{(Color online) a) Theoretical curves of the number of ions produced in 5 ms as a function of $U/\hbar$ solving equation (\ref{dGPE}) for different values of $w$. The values of $U_M$ obtained using the approximate expression given in the text are shown as open diamonds over the corresponding curves. b) Number of ions measured after 5 ms of dissipation for a thermal cloud as a function of the EB current, with $w=170 (7)$ nm. The solid line is the result of the corresponding numerical simulation using the molecular dynamics method. c) Comparison between the theoretical curves of the number of produced ions as a function of $I$ for the BEC and the \emph{corresponding} (see text) classical analogue ($w=170 (7)$ nm).}
\label{th}
\end{figure} 
 
\paragraph*{Dissipative dynamics} 
In order to gain a deeper insight into the dissipation-induced dynamics, we now look in detail at the time-resolved signals coming from the ion detector, reported in Fig. 4a. Initially the number of ions produced is well described by the exponential decay $\exp(-t\bar{\gamma})$, $\bar{\gamma}$ being the effective dissipation rate \cite{supp}. In Fig. 4b we show the integral of the signals in the first 5 $\mu$s, together with the simulated values as a function of the EB current. In this phase, where no paradoxical behavior is either observed or expected, the EB burns a hole in the BEC wavefunction (see Fig. 4d), defining a clear border between the space "inside" the hole and that "outside". Thereafter, the number of ions produced becomes almost constant, signaling the onset of a quasi-stationary dynamics \cite{zezu}. In this second phase, the reaction of the quantum system to the external perturbation takes place. When the strength of the dissipation is increased, the "outside" wavefunction passes from a situation of almost total transmission to a situation where reflection takes the leading role.  The non-monotonic dependence on the dissipation strength then becomes apparent (Fig. 4c). This represents the first experimental observation of the so-called back-flow paradox \cite{muga2}, i.e., of the onset of a temporary reflection from a localized perfect absorber. The curves plotted in Fig. 2 and 3 are then the sum of different contributions like those in Fig. 4b and 4c. % and this explains why the Fig. 2 and 3 curves do not approach zero for $I\rightarrow\infty$.

\begin{figure*}
	\centering
		\includegraphics[width=8.7cm]{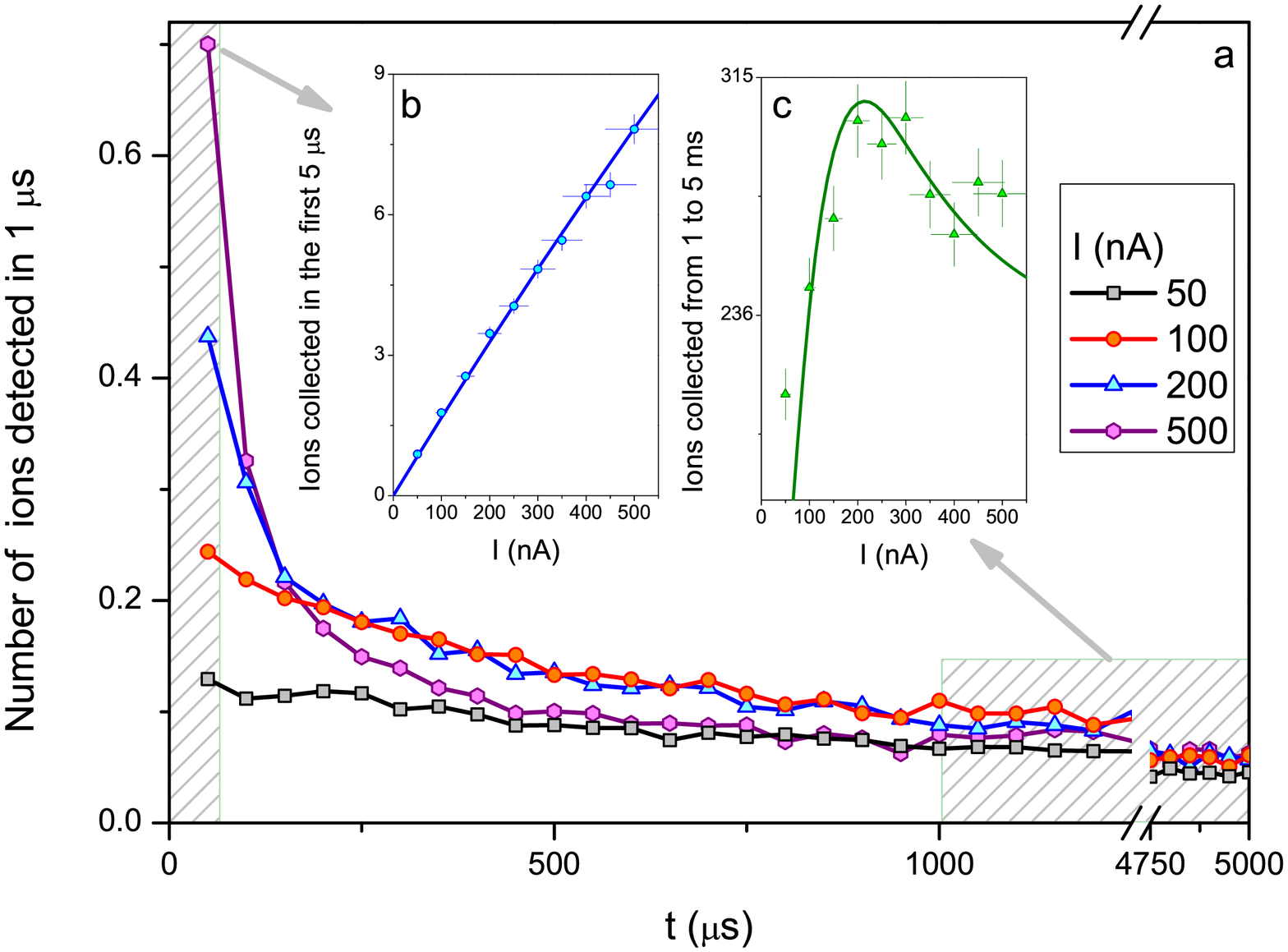}
		\includegraphics[width=8.4cm]{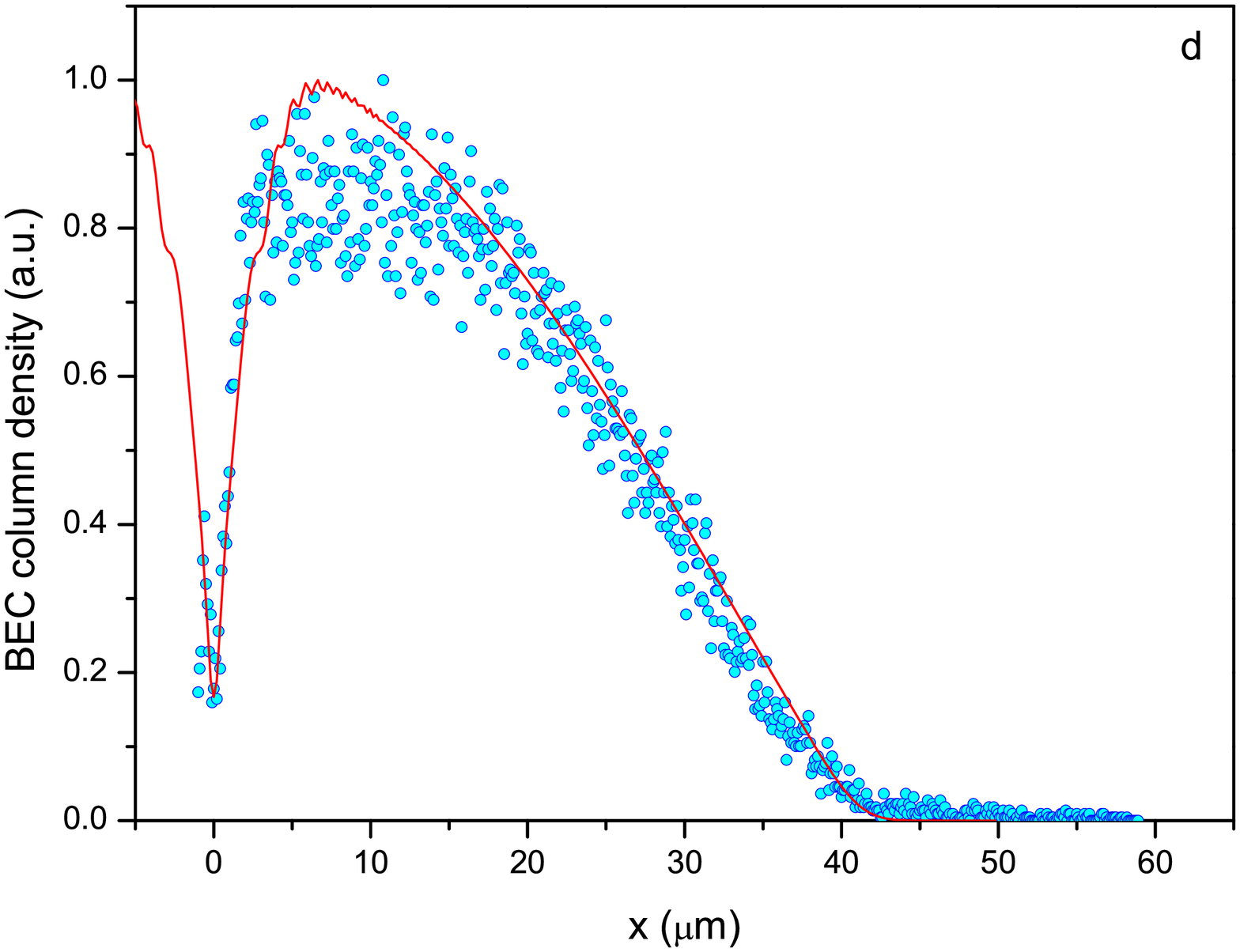}
	\caption{(Color online) a) Temporal resolved signal of the arrival time of the ions on the ion detector for different values of the EB current $I$ for $w=170(7)$ nm. In order to enhance readability the data have been plotted with a binning of 50 $\mu$s. The insets b) and c) show the integrals of the signal in the shaded areas (the first 5 $\mu$s and from 1 to 5 ms respectively) for different values of $I$ together with the theoretical calculations obtained solving equation (\ref{dGPE}). d) Points: scanning electron microscopy image of the BEC profile along the weak confining axis of the optical trap. The profile is the integrated column density along the direction of propagation of the EB. The scan is made after 1 ms of dissipation. The EB parameters are $I=150$ nA and $w=106(5)$ nm. The depletion of the density is visible in the origin, i.e., in the center of the BEC. The solid line is the profile obtained numerically solving equation (\ref{dGPE}).}
\label{t}
\end{figure*}

\paragraph*{Dissipation as continuous measurement}
Finally we demonstrate that the controlled dissipation is equivalent to a local measurement of the BEC density, i.e., of the squared modulus of its wavefunction. Starting from equation (\ref{dGPE}) and defining $\phi=\psi/\sqrt{N}$, where $N$ is the number of atoms, after some algebra we obtain the equation 
\begin{equation}
\frac{dN(t)}{dt}=-N(t)\int\gamma(\textbf{x})|\phi(\textbf{x},t)|^2d\textbf{x}.
\label{eq2}
\end{equation}  
The number of ions produced in a time interval $\Delta t$ around a certain time $t$ is $\Delta N_i(t)=\eta\int_{t-\Delta t/2}^{t+\Delta t /2}|dN(t)/dt| dt $. Hence we can conclude that what we perform is a direct measurement of the BEC density $|\psi(t)|^2$ in the region illuminated by the EB, as in \cite{gericke,Peter}. Since the seminal trilogy on the time of arrival in quantum mechanics \cite{allcock}, imaginary potentials have been linked to the action of a measurement apparatus while later refinements \cite{muga} formally demonstrated the equivalence between a pulsed measurement with period $\delta t$ and a continuous dissipative potential $U$, provided that $\delta t\simeq\hbar/U$. Our findings represent the experimental verification of the equivalence between the action of an imaginary potential and the one of a measurement apparatus on a quantum system. Indeed we show that a continuous measurement of the BEC density is nicely reproduced by introducing an imaginary potential in the corresponding Schr\"odinger equation.
Furthermore it is well understood theoretically \cite{misra,kuritzki}, and verified experimentally \cite{wineland,raizen,ketterle}, that performing continuous measurements strongly modifies the pre-existing dynamics of a quantum system. This is known as the Zeno (or inverse-Zeno) effect. In general, measurements are mainly performed on non-decaying systems and an extension of the standard definition to such systems is needed. In our case no pre-existing dynamics is present, since in the absence of the EB, the BEC is at rest. We have shown that the action of the continuous dissipative potential, or of the continuous measurement, strongly modifies the output of the measurement itself. In analogy with the standard definition we define \emph{dissipation induced Zeno dynamics} (DZD) when $dN_i(t)/dU<0$, $N_i(t)$ being the number of ions produced in the time $t$, and \emph{dissipation induced simple dynamics} (DSD) when $dN_i(t)/dU>0$. These definitions appear to be the natural extension of the standard ones, since the onset of the DZD requires large values of $U=\hbar\gamma(\textbf{0})/2$, which corresponds to pulsed measurements with small $\delta t$ \cite{muga}. From the definitions it follows that the DSD is observed where the dynamics is dominated by the Hamiltonian term of the Lindblad master equation, while the onset of the DZD corresponds to a dynamics governed by the dissipative term. We note that an effect resembling the DZD has been observed also in a system of decaying molecules in 1d \cite{rempe} and in an attractive Mott-insulator state \cite{hcn}.   

\paragraph*{Conclusion and Outlook}
We have experimentally demonstrated the implementation of an open many-body quantum system whose  Hamiltonian and dissipative dynamics can be independently and accurately controlled.  In the case of extremely strong and localized dissipation this can lead to the creation of dissipation-resistant states. The possibility to create such states in a controlled fashion can give new insights for engineering generalized environmental dark states. These kind of states are of fundamental interest and can possibly have practical applications in quantum computation schemes \cite{verstraete}. And inasmuch as our technique exploits the demonstrated link between dissipation and measurement, it can be used to address fundamental issues in quantum mechanics, like the definition of the time of arrival \cite{allcock}. The dissipation mechanism studied in the present paper is also particularly suited for lattice systems \cite{Kollath2}, thanks to its localized character and hence to the ability to selectively control the dissipation in a single lattice site. Indeed the use of the EB offers the unique possibility to create and study long-living exotic states in optical lattices \cite{sandro} and to characterize the interplay between dissipation and interactions \cite{supp,kollath}, and so would give access to the engineering of quantum phases in open quantum systems \cite{diehl}. 

\paragraph*{Acknowledgements}
We thank J.R. Anglin and A. Widera for reading the manuscript and for fruitful discussions. We are grateful to M. Scholl and P. W\"urtz for their help in the early stage of the experiment. We also thank V. Konotop and S. Wimberger for enlightening discussions. We acknowledge financial support by the DFG within the SFB/TRR 49 and GRK 792. G. B. and V. G. are supported by Marie Curie Intra-European Fellowships. R. L. acknowledges support by the MAINZ graduate school.

\section*{SUPPLEMENTARY MATERIALS}

\paragraph*{From the Lindblad Master Equation to the Gross-Pitaevskii equation}
We begin with a master equation of the Lindblad form \cite{book}
\begin{equation}
	i\hbar \dot{\hat{\rho}}=\left[ \hat{H},\hat{\rho} \right]+i\hbar\hat{\mathcal{L}}\hat{\rho},
\end{equation}
which describes the time evolution of the density operator of the system under the combined action of the Hamiltonian 
\begin{eqnarray}
\hat{H}=\int d^3x \hat{\Psi}^{\dagger}(\mathrm{\bf x})\left(-\frac{\hbar^2}{2m}{\nabla}^2+V_\mathrm{ext}(\mathrm{\bf x})\right)\hat{\Psi}(\mathrm{\bf x})+ \\ \nonumber
\frac{g}{2}\int d^3x \hat{\Psi}^{\dagger}(\mathrm{\bf x})\hat{\Psi}^{\dagger}(\mathrm{\bf x})\hat{\Psi}(\mathrm{\bf x})\hat{\Psi}(\mathrm{\bf x})
\end{eqnarray}
and a local loss term, described by the dissipation operator
\begin{eqnarray}
\hat{\mathcal{L}}\hat{\rho}=-\int d^3x \frac{\gamma(\mathrm{\textbf{x}})}{2}\left(\hat{\Psi}^{\dagger}(\mathrm{\bf x})\hat{\Psi}(\mathrm{\bf x})\hat{\rho}+ \right. \nonumber \\
\left.\hat{\rho} \hat{\Psi}^{\dagger}(\mathrm{\bf x})\hat{\Psi}(\mathrm{\bf x})-2\hat{\Psi}(\mathrm{\bf x})\hat{\rho}\hat{\Psi}^{\dagger}(\mathrm{\bf x})\right).
\end{eqnarray}
Here, $\hat{\Psi}^{\dagger}(\mathrm{\bf x})$ and $\hat{\Psi}(\mathrm{\bf x})$ are the bosonic field operators, obeying the bosonic commutation relation
\begin{equation}
\left[ \hat{\Psi}(\mathrm{\bf x}),\hat{\Psi}^{\dagger}(\mathrm{\bf x'})\right]=\delta^3(\mathrm{\bf x}-\mathrm{\bf x'}),
\end{equation}
the interaction strength is denoted with $g$ and $V_{\mathrm{ext}}(\mathrm{\bf x})$ is an external potential. The loss term $\gamma(\mathrm{\bf x})$ describes the rate of particle losses at the position $\mathrm{\bf x}$. Similar master equations for optical lattices with leaky sites have been studied in \cite{sandro,kollath}. The time evolution of an operator is then given by the differential equation 
\begin{equation}
\dot{\left\langle \hat{\Psi}(\mathrm{\bf x}) \right\rangle} = Tr\left( \hat{\Psi}(\mathrm{\bf x}) \dot{\hat{\rho}}\right ).
\end{equation}
Plugging equation (3) into (7) yields after some algebra
\begin{eqnarray}
i \hbar \dot{\left\langle \hat{\Psi}(\mathrm{\bf x}) \right\rangle} =
\left\langle \left[ -\frac{\hbar^2}{2m}{\nabla}^2+V_\mathrm{ext}(\mathrm{\bf x})+ \right. \right. \nonumber \\
\left.\left. g \hat{\Psi}^{\dagger}(\mathrm{\bf x}) \hat{\Psi}(\mathrm{\bf x})\right] \hat{\Psi}(\mathrm{\bf x})   \right\rangle - \frac{i\hbar\gamma(\mathrm{ \bf x})}{2}\left\langle \hat{\Psi}(\mathrm{\bf x}) \right\rangle.
\end{eqnarray}
Performing the usual mean-field approximation
\begin{equation}
\hat{\Psi}\simeq\hat{\Psi}^{\dagger}\simeq\langle\hat\Psi\rangle=\psi
\end{equation}
we arrive at equation (1) of the main text, i.e., to the time-dependent Gross-Pitaevski equation with imaginary potential (IGPE). The comparison between the experiment and the numerical solution of the IGPE suggests that the mean-field approximation is well justified for our parameters.

\paragraph*{Beam characterization}

We experimentally determine the value of $\sigma$ by monitoring the short-time decay of the ion signal for a fixed current and different electron beam sizes. Indeed analysing the solutions of the IGPE we have found that such a decay is approximately exponential with a time constant $\tau=1/\overline{\gamma}\simeq16ew^2/I\sigma$. The EB size $w$ is derived in two different ways. We first measure $w$ by fitting with an integrated gaussian function the electron signal originating from the scan of the beam through the edge of an auxiliary metallic target. The position of the beam focus and hence the value of $w$ at a fixed point can be easily varied by changing the current in its final focussing stage. A calibration of the beam size as a function of this current is also used to recover the measured $w$ when the working position of the microscope is finely adjusted at the actual location of the atoms. For comparison we also extract $w$ directly using the atomic sample as a target, by measuring the visibility of a two-dimensional deep optical lattice imaged by the EB itself. The lattice that we employed has a spacing of $547$ nm and a depth of $27E_r$. The beam size $w$ is obtained by deconvolving the density distribution of the atoms in the lattice by their expected Wannier functions distribution. Note that the EB size does not vary appreciably along the radial extension of the cloud ($\sim 7$ $\mu$m), as its Rayleigh length is always conveniently larger ($>30$ $\mu$m). The two measurements yield compatible results and are also in accordance with the measurement of $1/\tau$ as a function of the current $I$ performed for two different beam sizes: $w=170(7)$ nm and $w=255(10)$ nm, see Fig. S1. From this analysis we obtain the value of the electron-BEC cross-section used in this work: $\sigma=3.6(0.1)\times10^{-21}$ m$^2$. 

\begin{figure}
\renewcommand{\thefigure}{S1}
\begin{center}
\includegraphics[width=8.5cm]{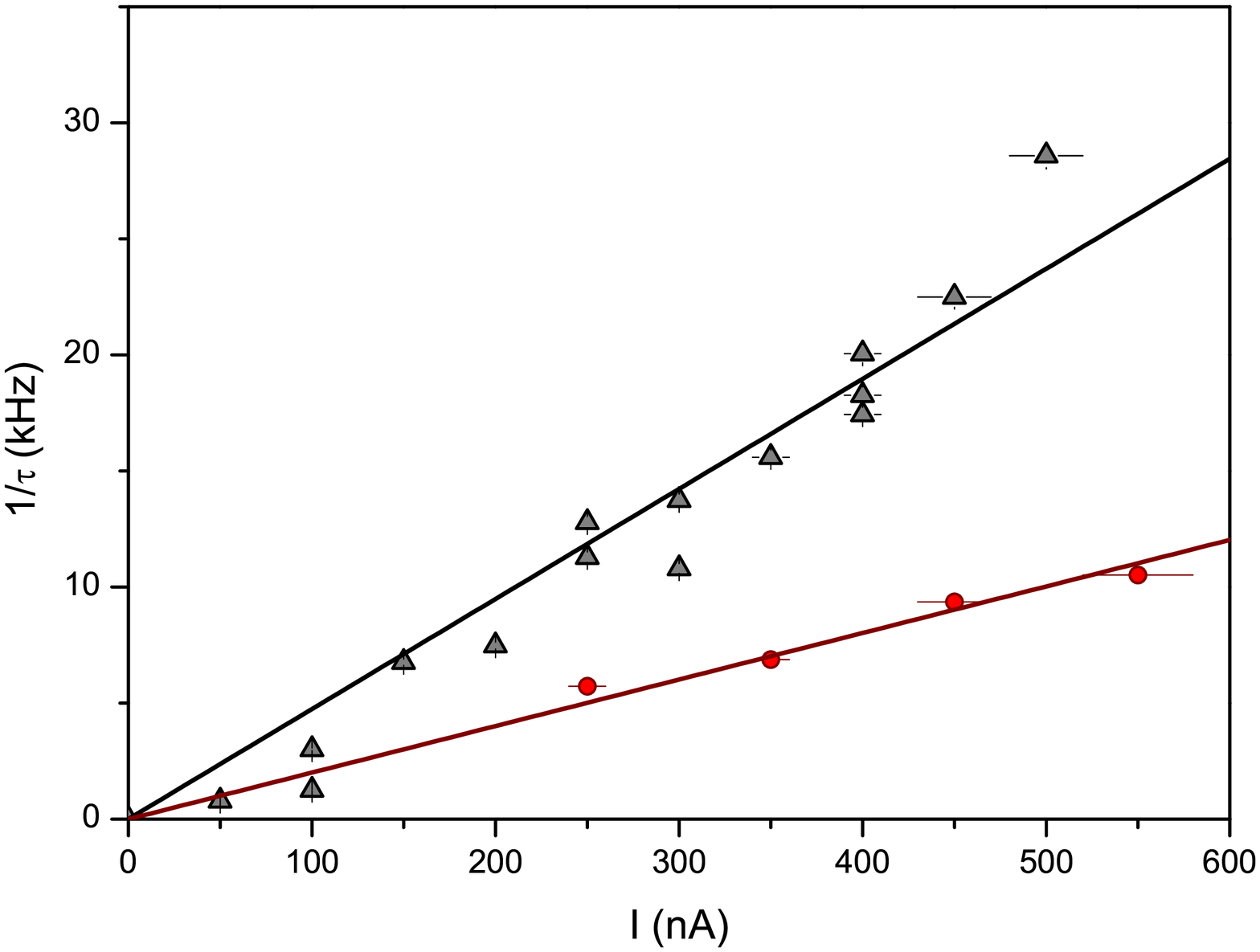}
\end{center}
\caption{(Color online) Inverse decay time $1/\tau$ as a function of the EB current $I$ for two different values of the beam size $w=170 (7)$ nm (gray triangles) and $w=255 (10)$ nm (red circles). The solid lines are linear fits to the data. The current $I$ is measured by means of a Faraday cup, which is integrated in the experimental apparatus.} 
\label{fig:taui}
\end{figure}
                  
\paragraph*{Secondary Effects}

The main spurious effect is caused by the losses due to the secondary collisions of the produced ions with the trapped atoms. Indeed such ions, produced in the center of the BEC, travelling through the atomic cloud can collide with other atoms expelling them from the trap. This effect can in principle be reduced applying a strong electric field which extracts the ions faster; but it cannot be completely avoided. In order to take into account this loss channel the IGPE must be modified adding an ion-atom collision term in the imaginary potential such that:
\begin{equation}
\frac{\hbar\gamma(\textbf{x})}{2}\rightarrow\frac{\hbar\gamma(\textbf{x})}{2}+\frac{\hbar\alpha(\textbf{x})}{2}\frac{dN_i(t)}{dt}
\end{equation} 
where $dN_i(t)/dt$ is the ions production rate and $\alpha(\textbf{x})=A\exp(-(x^2+y^2)/2W^2)$ is a gaussian function whose amplitude $A$ and extension $W$ depend on the extraction geometry of the ions. We have determined these two parameters fitting the time resolved curves like the one in Fig. 1b, finding $A=1\times10^{-3}$ and $W=20w$. Taking into account this effect is important in our case mainly to estimate our ion detection efficiency, that is $\simeq75\%$, but it has no significant impact on the physics presented. In the main text when we compare the theory with data we have included this effect while pure theoretical graphs report the ideal case. 

Additional spurious effects that, in principle, must be taken into account are related to the real potentials stemming from the EB. Indicating with $R$ the radial distance from the center of the EB in the direction perpendicular to its propagation, one can find that the EB generates a magnetic field which has the following expression:
\begin{equation}
B_e(R)=\frac{\mu_0IR}{4\pi w^2}e^{-\frac{R^2}{2w^2}}.
\end{equation}
The electron microscope machine produces itself a background magnetic field $B_{bg}$ of $\simeq$ 1 G, directed along the direction of propagation of the EB and is always present at the atoms position. Due to this, the total magnetic field acting on the atoms is $B(\textbf{x})=\sqrt{B_e^2+B_{bg}^2}$. Since $B_e$ is on the order of just 1-2 mG, it results that $B\simeq B_{bg}$, making any effect due to the magnetic field produced by the EB $B_e$ negligible.

We now consider the electric field induced on the atoms by the flux of electrons, which can be approximated by:
\begin{equation}
E_e(R)=\frac{I}{2\pi\epsilon_0Rv_e}(1-e^{-\frac{R^2}{2w^2}})
\end{equation}
where $v_e$ is the mean velocity of the electrons.
This field leads to an effective potential orientated along the direction perpendicular to the beam propagation $U_E(R)=-1/2\alpha_0E_e^2(R)$ where $\alpha_0$ is the ground state polarizability of the $^{87}$Rb atoms. For the beam maximum current we employed in the present experiments, $I=500$ nA, the ratio between the maximum modulus of the dissipative and electric potentials is $\simeq580$ and this latter contribution is again negligible. Finally, we have also taken into account for the effects of the real potentials of the EB in the numerical simulations finding, as expected, no appreciable correction to the calculations performed without.

\paragraph*{Molecular Dynamics With Dissipation}

The numerical simulations made to reproduce the data for the thermal cloud were performed using a molecular dynamics method. We have used a Runge-Kutta algorithm to simulate the dynamics of $4\times$10$^5$ atoms at the temperature of 1 $\mu$K in a harmonic potential. For this simulation the atoms were supposed to be point-like, neglecting any kind of quantum effect like the finite extension due to the deBroglie wavelength. Moreover we did not take into account the interactions between the atoms. The action of the electron beam was simulated introducing $I/e\Delta t$ electrons per m$^2$, gaussian distributed in space along $x$ and $y$, and uniformly distributed along $z$, the direction of propagation of the EB. $\Delta t$ is the time step of the simulation. Every time the distance $r$ between one atom and one electron is smaller than $\sqrt{\sigma/\pi}$ the atom is supposed to collide with the electron leading to the loss of such an atom from the trap. We have found that this method, despite its simplicity, reproduces quite well our data except for a multiplicative factor, which takes into account the efficiency of the ion detector and the branching-ratio of the collision processes.

\paragraph*{Effect of the interactions}

We point out that a big advantage originating from the many-body nature of the BEC is the possibility to perform continuous measurements on a single wavefunction, giving direct access to the dynamics without the need of ensemble averages. At the same time the interactions between the atoms play a crucial role in determining the properties of such a wavefunction, through the non-linear term in the IGPE, and have a direct impact on the dissipative dynamics. Even if experimentally we are not able to change the interactions we can do it when solving the IGPE changing the value of $a$. The results are reported in Fig. S2. 

\begin{figure}
\renewcommand{\thefigure}{S2}
\begin{center}
\includegraphics[width=8.5cm]{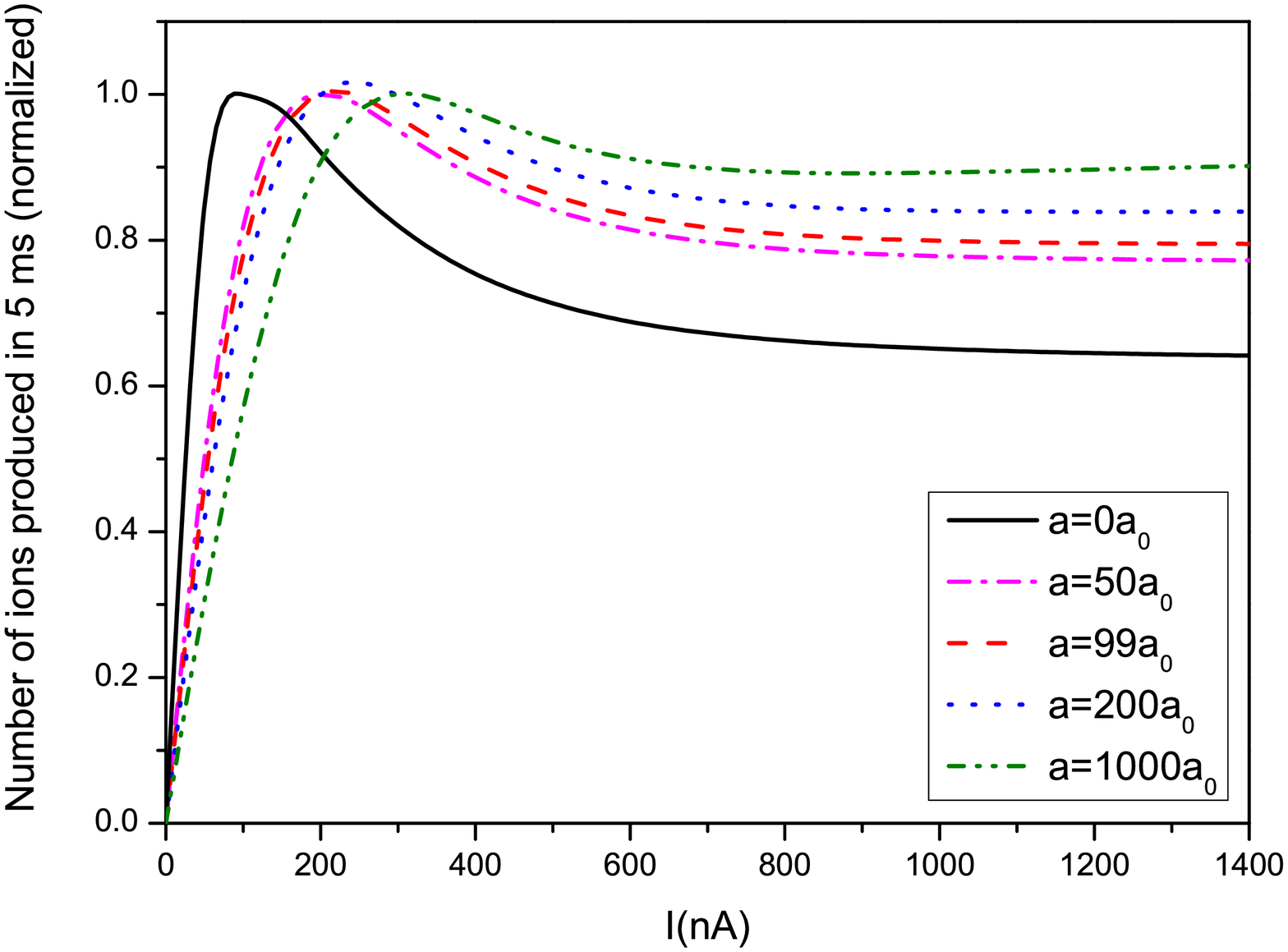}
\end{center}
\caption{(Color online) Calculated total number of ions produced in the first 5 ms as a function of the EB current for different values of the scattering length $a$. The position of the maximum dissipation is shifted to higher values of $I$ as the interactions between the atoms are increased. The normalization is made to compensate the effect of the changing density ($\propto a^{-3/5}$). Clearly increasing interactions hinder the onset and the effects of the DZD.}
\label{interactions}
\end{figure}

As expecting a change in the interaction strength changes the position of the maximum of dissipation, since it depends on $\mu$: the stronger is the repulsion between the atoms the more the onset of the DZD is shifted to larger values of $I$ and the more its effects are decreased. Moreover an increasing in $a$ implies a reduced production of ions, for a given $\gamma$, due to the subsequent change in the density. As shown in Fig. S2, also a non-interacting BEC displays the non monotonic behaviour observed in the experiment. From this we hence conclude that the observed effects originate from the wave nature of the BEC and not from its many-body character.  Nevertheless, according to Fig. S2, the mean-field potential stemming from the atomic interactions represents another important controllable parameter that can be used to engineer the response of the system to the environmental action. 

\end{document}